\documentclass[12pt]{article}
\usepackage{amsfonts}
\newcommand{\beqn}{\begin{eqnarray}}
\newcommand{\eeqn}{\end{eqnarray}}
\newtheorem{theorem}{Theorem}

\newtheorem{lemma}{Lemma}

\newcommand{\proof}{\noindent {\it Proof.\/}\ }
\newcommand{\qed}{{\it Q.E.D.\/} \bigskip\par}
\newcommand{\rd}{\partial}

\newcommand{\diag}{\mathop{{\mathrm{diag}}}}
\newcommand{\Tr}{\mathop{{\mathrm{Tr}}}}
\newcommand{\rank}{\mathop{{\mathrm{rank}}}}
\newcommand{\mod}{\mathop{{\mathrm{mod}}}}
\newcommand{\const}{{\mathrm{const.}}}
\newcommand{\id}{{\mathrm{id}}}
\newcommand{\Jac}{{\mathrm{Jac}}}
\newcommand{\Prym}{{\mathrm{Prym}}}
\newcommand{\SW}{{\mathrm{SW}}}
\newcommand{\KdV}{{\mathrm{KdV}}}
\newcommand{\DW}{{\mathrm{DW}}}
\newcommand{\bbC}{{\mathbb{C}}}
\newcommand{\bbP}{{\mathbb{P}}}
\newcommand{\bbZ}{{\mathbb{Z}}}
\newcommand{\frakg}{{\mathfrak{g}}}
\newcommand{\calA}{{\mathcal{A}}}
\newcommand{\calB}{{\mathcal{B}}}
\newcommand{\calC}{{\mathcal{C}}}
\newcommand{\calD}{{\mathcal{D}}}
\newcommand{\calF}{{\mathcal{F}}}
\newcommand{\calN}{{\mathcal{N}}}
\newcommand{\calO}{{\mathcal{O}}}
\newcommand{\calP}{{\mathcal{P}}}

\newcommand{\calU}{{\mathcal{U}}}
\newcommand{\calX}{{\mathcal{X}}}

\newcommand{\Xtilde}{\tilde{X}}
\newcommand{\Ztilde}{\tilde{Z}}
\newcommand{\Omegahat}{\hat{\Omega}}
\newcommand{\abar}{\bar{a}}
\begin{document}

\title{Whitham Deformations of Seiberg-Witten Curves\\
for Classical Gauge Groups}
\author{Kanehisa Takasaki\\
{\normalsize Department of Fundamental Sciences, Kyoto University}\\
{\normalsize Yoshida, Sakyo-ku, Kyoto 606-8501, Japan}\\
{\normalsize E-mail: takasaki@yukawa.kyoto-u.ac.jp}}
\date{}
\maketitle

\vfill

\begin{abstract} 
Gorsky et al. presented an explicit construction of Whitham 
deformations of the Seiberg-Witten curve for the $SU(N+1)$ 
$\calN = 2$ SUSY Yang-Mills theory.  We extend their result 
to all classical gauge groups and some other cases such as 
the spectral curve of the $A^{(2)}_{2N}$ affine Toda  
system.  Our construction, too, uses fractional powers of 
the superpotential $W(x)$ that characterizes the curve.  
We also consider the $u$-plane integral of topologically 
twisted theories on four-dimensional manifolds $X$ with 
$b_2^{+}(X) = 1$ in the language of these explicitly 
constructed Whitham deformations and an integrable hierarchy 
of the KdV type hidden behind.  
\end{abstract}
\bigskip

\begin{flushleft}
{\tt KUCP-0127}\\
{\tt hep-th/9901120}
\end{flushleft}

%%%%%%%%%%%%%%%%%%%%%%%%%%%%%%%%%%%%%%%%%%%%%%%%%%%%%%%%%%%%%%%%%%%%
\newpage
\renewcommand{\theequation}{\arabic{section}.\arabic{equation}}

\section{Introduction}
\setcounter{equation}{0}

The geometry of Seiberg and Witten's low energy effective 
theory of the four-dimensional $\calN = 2$ SUSY Yang-Mills 
theories (with and without matters) \cite{bib:Se-Wi} are 
based on complex algebraic curves now generally called 
the ``Seiberg-Witten curves''.  Coordinates of the moduli 
space (Coulomb branch) $\calU$ of these curves are given 
by the Casimirs $u_j$ ($j = 1,\cdots,N = \rank(G)$) of 
the scalar field $\phi$ of the $\calN = 2$ vector multiplet.  
Each curve carries a special meromorphic differential 
(the ``Seiberg-Witten differential'') $dS_{\SW}$.  
Given a suitable set of cycles $A_j,B_j$ ($j=1,\cdots,N$), 
this differential $dS_{\SW}$ induces a ``special geometry'' 
on the moduli space $\calU$.  The ``special coordinates''
are given by the period integrals $a_j$ and $a_j^D$ of 
$dS_{\SW}$ along these cycles.  The prepotential $\calF$ 
of the low energy effective theory is determined by this 
special geometry. 

Gorsky et al. \cite{bib:GKMMM} and Martinec and Warner 
\cite{bib:Ma-Wa} discovered that an integrable system, 
typically an affine Tida system, is hidden behind this 
geometric setup.  This point of view soon turned out 
to be very useful not only for studying various 
four-dimensional $\calN = 2$ SUSY gauge theories 
\cite{bib:Na-Ta,bib:Do-Wi,bib:Eg-Ya-RG,bib:Martinec,%
bib:Go-Ma,bib:It-Mo,bib:Marshakov,bib:Ah-Na,bib:Kr-Ph,%
bib:DH-Kr-Ph}
but also for elucidating a universal mathematical 
structure underlying those examples 
\cite{bib:Donagi,bib:Freed}.  
In fact, all the building blocks of the Seiberg-Witten 
theory fit to the language of integrable systems. 
The Seiberg-Witten curve is nothing but the spectral 
curve of the integrable system.  The differential 
$dS_{\SW}$ is related to its complex symplectic 
structure.  The special coordinates $a_j$ are 
identified with the action variables; the angle 
variables live on an Abelian subvariety (called 
the ``special'' or ``distinguished'' Prym variety) 
of the Jacobi variety of the Seiberg-Witten curve.  
The Casimirs $u_j$ give an involutive set of 
Hamiltonians.  The commuting flows generated by 
these Hamiltonians sweep out the aforementioned 
Prym variety. In a more geometric language, the 
phase space $\calX$ has a Lagrangian fibration 
$\calX \to \calU$ by these $N$-dimensional complex 
``Lieouville tori''. 

This link with integrable systems implies the existence 
of another kind of deformations of the Seiberg-Witten 
curves: Whitham deformations \cite{%
bib:GKMMM,bib:Ma-Wa,bib:Na-Ta,bib:It-Mo,bib:Marshakov}
The differential equations that characterize these 
deformations, are called ``Whitham equations''. 
These deformations are parametrized by an extra set of 
variables $T_n$ ($n=1,2,\cdots$). These variables are 
referred to as ``slow variables'' in the theory of 
Whitham equations; ``fast variables'' $t_n$ are the 
time variables of commuting Hamiltonian flows in the 
integrable system (or an integrable hierarchy).  

Several interesting physical interpretations of these 
Whitham deformations have been proposed.  Deformations 
by $T_1$ are identified to be the renormalization group 
flows \cite{bib:Matone,bib:So-Th-Ya,bib:Eg-Ya-RG,%
bib:Bo-Ma,bib:DH-Kr-Ph}.
In this sense, the other Whitham deformations may be 
thought of as generalized RG flows.  Gorsky et al. 
\cite{bib:GMMM} constructed an explicit solution of 
the Whitham equations for the $SU(N+1)$ Seiberg-Witten 
curve, and argued its relation to topologically twisted 
gauge theories.  Edelstein et al. \cite{bib:Ed-Ma-Ma} 
interpreted the Whitham deformations of Gorsky et al. 
as soft breaking of $\calN = 2$ SUSY by spurion fields. 

In this paper we generalize the construction of Gorsky 
et al. \cite{bib:GMMM} to other classical gauge groups.  
We consider the $\calN = 2$ SUSY Yang-Mills theories 
in the vector representation of classical gauge groups. 
As Martinec and Warner pointed out \cite{bib:Ma-Wa}, 
the Seiberg-Witten curves for these theories can be 
written in a unified form, see (\ref{eq:SW-curve}), 
which is essentially the spectral curves of affine 
Toda systems. This expression contains a function 
$W(x)$ (called the ``superpotential'' in analogy 
with topological Landau-Ginzburg theories of A-D-E 
singularities).  In the case of the $SU(N+1)$ 
Seiberg-Witten curve, $W(x)$ is a polynomial.  
Gorsky et al. use fractional powers of $W(x)$ very 
ingeniously.  Actually, their method work for 
a general rational superpotential. We demonstrate 
it in the case of the Seiberg-Witten curves for 
other classical gauge groups and some other Toda 
spectral curves.  

We also address another issue that is raised in our 
previous paper \cite{bib:Ta-u-plane}. As discussed therein, 
an interesting interplay of the ``slow variables'' $T_n$ 
and the ``fast variables'' $t_n$ can be observed in the 
$u$-plane integral of the topologically twisted theories 
on four-dimensional manifolds $X$ with $b_2^{+}(X) = 1$ 
\cite{bib:Mo-Wi,bib:Ma-Mo,bib:Lo-Ne-Sh}. ($b_2^{+}(X)$ 
is the self-dual part of the 2nd Betti number.) 
Our consideration in the previous paper was limited 
to the case of $SU(N+1)$.  We revisit this issue, 
now armed with the explicit construction of Whitham 
deformations for all classical gauge groups.  

This paper is organized as follows.  Section 2 is 
a collection of basic mathematical notions concerning 
Seiberg-Witten curves.  Particularly important are 
the Prym varieties and related differentials that are 
crucial in handling the cases other than $SU(N+1)$. 
Section 3 is a review of the construction of 
Gorsky et al.  We present all technical details, which 
are used in the subsequent sections.  In Section 4 
we consider the case of the $SO(2N)$ Seiberg-Witten 
curve in detail.  Section 5 deals with the case of 
$SO(2N+1)$ and $Sp(2N)$, along with some other cases 
that are not directly related to $\calN = 2$ SUSY 
Yang-Mills theories but can be treated similarly.  
In Section 6 we consider the $u$-plane integral of 
the topologically twisted theories.  Section 7 is 
devoted to discussions.  Appendix is added to show 
a precise form of the spectral curves of the affine 
Toda systems in the usual Lax formalism.

%%%%%%%%%%%%%%%%%%%%%%%%%%%%%%%%%%%%%%%%%%%%%%%%%%%%%%%%%%%%%%%%%%%%

\section{Curves, Differentials, and Prym Varieties}
\setcounter{equation}{0}

\subsection{Various complex algebraic curves}

The Seiberg-Witten curves of $\calN = 2$ SUSY Yang-Mills 
theories in the vector representation of classical 
gauge groups can be written in the following common 
form \cite{bib:Ma-Wa}: 
\beqn
   z + \frac{\mu^2}{z} = W(x). 
   \label{eq:SW-curve}
\eeqn
Here $\mu$ is some power of the renormalization group 
parameter $\Lambda$, and $W(x)$ (the ``superpotential'') 
a polynomial or a Laurent polynomial of the following form:
\beqn
    SU(N+1) &:&  W(x) = x^{N+1} - \sum_{j=2}^{N+1} u_j x^{N+1-j}. 
    \nonumber \\
    SO(2N+1) &:& W(x) = x^{-1} \left(x^{2N} 
      - \sum_{j=1}^N u_j x^{2N-2j} \right). 
    \nonumber \\
    Sp(2N) &:&  W(x) = x^2 \left(x^{2N} 
      - \sum_{j=1}^N u_j x^{2N-2j} \right) + 2\mu. 
    \nonumber \\
    SO(2N) &:& W(x) = x^{-2} \left(x^{2N} 
      - \sum_{j=1}^N u_j x^{2N-2j} \right). 
    \nonumber 
\eeqn
The polynomials 
\beqn
    P(x) = x^{N+1} - \sum_{j=2}^{N+1} u_j x^{N+1-j}
\eeqn
for the $SU(N+1)$ gauge group and 
\beqn
    Q(x^2) = x^{2N} - \sum_{j=1}^N u_j x^{2N-2j} 
\eeqn
for the other gauge groups may be identified with 
the characteristic polynomial $\det(xI - \phi)$ 
of the scalar field $\phi$ of the $\calN = 2$ SUSY 
vector multiplet.  The coefficients $u_j$ are 
accordingly the expectation value of the Casimirs 
of $\phi$.  In the case of $SU(N+1)$ and $SO(2N)$, 
$W(x)$ coincides with the superpotential of the 
topological Landau-Ginzburg theories (or $d < 1$ 
strings) of singularities of the A and D type 
\cite{bib:DVV,bib:Bl-Va}. Because of this, $W(x)$ 
is referred to as the ``superpotential''. 

As Martinec and Warner pointed out \cite{bib:Ma-Wa}, 
these curves coincide with the spectral curves of 
the affine Toda system of the type $\frakg^{(1)\vee}$ 
dual to the the untwisted affine Lie algebra 
$\frakg^{(1)}$ of the gauge group $G$.  In particular, 
the affine Toda systems for the non-simply-laced 
gauge groups $SO(2N+1)$ and $Sp(2N)$ are of the 
twisted affine type $B_N^{(1)\vee} = A_{2N-1}^{(2)}$ 
and $C_N^{(1)\vee} = D_{N+1}^{(2)}$.  
The affine Toda systems for the other classical 
affine Lie algebras, too, have spectral curves of 
the above form: 
\beqn
    \mbox{$B_N^{(1)},C_N^{(1)}$-Toda} &:&
      W(x) = x^{2N} - \sum_{j=1}^N u_j x^{2N-2j}, 
      \nonumber \\
    \mbox{$A_{2N}^{(2)}$-Toda} &:& 
      W(x) = x \left( x^{2N} - \sum_{j=1}^N u_j x^{2N-2j} \right). 
      \nonumber 
\eeqn

All these curves are hyperelliptic.  The following is 
an equivalent expression in the usual expression 
$y^2 = R(x)$ of hyperelliptic curves.  Actually, it is 
in this form (or a quotient curve discussed later on) 
that the Seiberg-Witten curves for classical gauge groups 
were first derived 
\cite{bib:Ar-Fa-SU,bib:Da-Su-SO,bib:Br-La-SO,bib:Ar-Sh-all}. 
\begin{enumerate}
\item The Seiberg-Witten curves: 
\beqn
    SU(N+1) &:& 
      y^2 = P(x)^2 - 4 \mu^2, \ 
      z = \bigl(P(x) + y \bigr) / 2. 
      \nonumber \\
    SO(2N+1) &:& 
      y^2 = Q(x^2)^2 - 4 \mu^2 x^2, \ 
      z = \bigl(Q(x^2) + y \bigr) / 2x. 
      \nonumber \\
    Sp(2N) &:& 
      y^2 = Q(x^2) \bigl(x^2 Q(x^2) + 4 \mu \bigr), \ 
      z = \bigl(2 \mu + x^2 Q(x^2) + xy \bigr) / 2. 
      \nonumber \\
    SO(2N) &:& 
      y^2 = Q(x^2)^2 - 4 \mu^2 x^4, \ 
      z = \bigl( Q(x^2) + y \bigr) / 2. 
      \nonumber 
\eeqn
\item Other affine Toda spectral curves: 
\beqn
    B_N^{(1)},C_N^{(1)} &:& 
      y^2 = Q(x^2)^2 - 4 \mu^2, \ 
      z = \bigl( Q(x^2) + y \bigr) / 2. 
      \nonumber \\
    A_{2N}^{(2)} &:& 
      y^2 = x^2 Q(x^2)^2 - 4 \mu^2, \ 
      z = \bigl( x Q(x^2) + y \bigr) / 2. 
      \nonumber 
\eeqn
\end{enumerate}
The curves other than the $SU(N+1)$ Seiberg-Witten curve 
can be classified into two groups: 
\begin{itemize}
\item A: 
The $Sp(2N)$ Seiberg-Witten curve and 
the $A_{2N}^{(2)}$-Toda curve. 
\item B: 
The other curves. 
\end{itemize}
We shall show that the curves in the two groups exhibit 
different properties in many aspects.  The first aspect 
that we now point out, is their genera: 
\begin{itemize}
\item Case A: The curve has genus $2N$.  
\item Case B: The curve has genus $2N - 1$.
\end{itemize}

\subsection{Involutions and Prym varieties}

The above curves, which we denote by $C$, have several 
involutions.  Common to all are the hyperelliptic involution 
\beqn
    \sigma_1: \quad 
      (x,y) \mapsto (x,-y), \quad  (x,z) \mapsto (x, \mu^2/z). 
\eeqn
All the curves other than the $SU(N+1)$ curves have 
the second involution 
\beqn
    \sigma_2: \quad 
      (x,y) \mapsto (-x,y), \quad (x,z) \mapsto (-x,z). 
\eeqn
The quotient $C_2 = C / \sigma_2$ by the second involution 
is also a hyperelliptic curve.  It has different properties 
in accordance with the above classification: 
\begin{itemize}
\item Case A: $C_2$ has genus $N$, and the covering map 
$C \to C_2$ is unramified. 
\item Case B: $C_2$ has genus $N-1$, and the covering map 
$C \to C_2$ is ramified. 
\end{itemize} 

The double covering $C \to C_2$ determines the Prym variety 
$\Prym(C/C_2)$.  This is an $N$-dimensional Abelian variety, 
which plays the role of the Jacobi variety $\Jac(C)$ for 
the $SU(N+1)$ Yang-Mills theory.  (Fay's book \cite{bib:Fay} 
provides us with useful information on this kind of Prym 
varieties.)  Following Fay's book, let us specify the 
structure of this Prym variety in more detail.   

The algebro-geometric definition of this Prym variety is 
based on an automorphism $\sigma_2: \Jac(C) \to \Jac(C)$ 
induced by the involution $\sigma_2$.  Consider the Jacobi 
variety as the set of the linear equivalence classes of 
divisors $\calD$ of degree zero. The Prym variety 
$\Prym(C/C_2)$, by definition, is the image of 
$\id - \sigma_2$ (where $\id$ is the identity map), 
namely, consists of the linear equivalence classes of 
divisors of the form $\calD - \sigma_2(\calD)$. 

An equivalent complex analytic expression is the complex 
torus 
\beqn
    \Prym(C/C_2) \simeq 
    \bbC^N / (\Delta \bbZ^N + 2 \calP \bbZ^N), 
\eeqn
where $\Delta$ is a diagonal matrix 
$\Delta = \diag(d_1,\cdots,d_N)$ with positive integers 
on the diagonal line, and $\calP$ is a complex symmetric 
matrix $(\calP_{jk})$ with positive definite imaginary part. 
This Prym variety is thus a polarized Abelian variety with 
the following polarization $(d_1,\cdots,d_N)$:
\begin{itemize}
\item Case A: $(d_1,\cdots,d_N) = (2,\cdots,2,2)$. 
\item Case B: $(d_1,\cdots,d_N) = (2,\cdots,2,1)$. 
\end{itemize}
In particular, Case A is substantially a principally polarized 
Abelian variety with period matrix $\calP$; the above expression 
is simply for dealing with the two cases in a unified way. 

The matrix elements of $\calP$ are period integrals of 
holomorphic differentials $d\omega_j$ ($j = 1,\cdots,N$) 
that are ``odd'' under the action of $\sigma_2$:
\beqn
    \sigma_2^* d\omega_j = - d\omega_j. 
\eeqn
These differentials are uniquely determined by the normalization 
condition 
\beqn
    \oint_{A_j} d\omega_k = \delta_{jk}, 
\eeqn
and the matrix elements $\calP_{jk}$ of $\calP$ are given by 
\beqn
    \calP_{jk} = \frac{d_j}{2} \oint_{B_j} d\omega_k. 
\eeqn
The $2N$ cycles $A_j,B_j$ ($j = 1,\cdots,N$) in these period 
integrals have to be chosen as follows: 
\begin{itemize}
\item Case A: 
The $4N$ cycles $A_j,-\sigma_2(A_j),B_j,-\sigma_2(B_j)$ 
($j = 1,\cdots,N$) form a symplectic basis of cycles 
of $C$. 
\item Case B: 
The homology classes $[A_N]$ and $[B_N]$ are ``odd'' under 
the action of $\sigma_2$, i.e., $\sigma_2([A_N]) = - [A_N]$ 
and $\sigma_2([B_N]) = - [B_N]$.  The $4N - 2$ cycles 
$A_j,-\sigma_2(A_j),B_j,-\sigma_2(B_j)$ ($j = 1,\cdots,N-1$) 
and $A_N,B_N$ altogether form a symplectic basis of cycles 
of $C$. 
\end{itemize}
In particular, these cycles have the intersection numbers 
$A_j \cdot A_k = B_j \cdot B_k = 0$ and 
$A_j \cdot B_k = \delta_{jk}$.

\subsection{Seiberg-Witten differential}

The Seiberg-Witten differential is given by 
\beqn
    dS_{\SW} = x \frac{dz}{z} 
             = \frac{xW'(x)dx}{\sqrt{W(x)^2 - 4\mu^2}}. 
\eeqn
The following list shows a more explicit form of this 
differential.  
\begin{enumerate}
\item For the Seiberg-Witten curves: 
\beqn
    SU(N+1) &:& 
      dS_{\SW} = xP'(x) \frac{dx}{y}. 
      \nonumber \\
    SO(2N+1) &:& 
      dS_{\SW} = \Bigl(2Q'(x^2)x^2 - Q(x^2)\Bigr)\frac{dx}{y}. 
      \nonumber \\
    Sp(2N) &:& 
      dS_{\SW} = x \Bigl(2Q'(x^2)x^2 + 2Q(x^2)\Bigr)\frac{dx}{y}. 
      \nonumber \\
    SO(2N) &:& 
      dS_{\SW} = \Bigl(2Q'(x^2)x^2 - 2Q(x^2)\Bigr)\frac{dx}{y}. 
      \nonumber 
\eeqn
\item For the other Toda curves: 
\beqn
    B_N^{(1)},C_N^{(1)} &:& 
      dS_{\SW} = 2Q'(x^2)x^2 \frac{dx}{y}. 
      \nonumber \\
    A_{2N}^{(2)} &:& 
      dS_{\SW} = x \Bigl(2Q'(x^2)x^2 + Q(x^2)\Bigr)\frac{dx}{y}. 
      \nonumber 
\eeqn
\end{enumerate}
Here the prime  means differentiating by $x$, i.e., 
$P'(x) = dP(x)/dx$, etc.  

A fundamental property of this differential is that 
it generates holomorphic differentials on $C$ as follows. 
Let $\left.(\rd/\rd u_j)\cdots\right|_{z=\const}$ 
denote differentiating the quantity inside by $u_j$ 
while keeping $z$ constant.  For the Seiberg-Witten 
differential, this gives 
\beqn
    \frac{\rd}{\rd u_j}\left.dS_{\SW} \right|_{z=\const} 
    = \left.\frac{\rd x}{\rd u_j}\right|_{z=\const} 
        \frac{dz}{z} 
    = - \frac{\rd W(x)/\rd u_j}{\rd W(x) /\rd x} 
          \frac{xW'(x)dx}{\sqrt{W(x)^2 - 4\mu^2}}. 
    \label{eq:dSSW-by-u}
\eeqn
Here we have also used the relation 
\beqn
    \frac{\rd W(x)}{\rd u_j} 
    + \frac{\rd W(x)}{\rd x} 
      \left.\frac{\rd x}{\rd u_j}\right|_{z=\const} 
    = 0, 
\eeqn
which follows from the equation of the curve $C$.  
One can verify, for each case presented above, that 
all these differentials for $j = 1,\cdots,N$ are 
holomorphic differentials.  In the case of the 
$SU(N+1)$ Seiberg-Witten curve, these $N$ holomorphic 
differentials form a basis of the space of holomorphic 
differentials on $C$.  In the other cases, these 
differentials are also linearly independent, 
but ``odd'' under the action of $\sigma_2$, because 
the Seiberg-Witten differential itself is also ``odd'': 
\beqn
    \sigma_2^* dS_{\SW} = - dS_{\SW}. 
\eeqn
The $u_j$-derivatives (\ref{eq:dSSW-by-u})  give 
a basis of ``odd'' holomorphic differentials (also called 
``Prym differentials''). 

Given a set of cycles $A_j$ and $B_j$ mentioned above, 
one can define the special coordinates $a_j$ and their 
duals $a_j^D$ on the moduli space $\calU$ as follows: 
\beqn
    a_j = \oint_{A_j} dS_{\SW}, \quad 
    a_j^D = \oint_{B_j} dS_{\SW}. 
\eeqn
The $N$ functions $a_j$ ($j = 1,\cdots,N$), as well as 
the $a_j^D$'s, are functionally independent and give 
a local coordinate system on $\calU$.  Differentiating 
the Seiberg-Witten differential now by $a_j$'s give 
the normalized holomorphic differentials $d\omega_j$: 
\beqn
    \left.\frac{\rd}{\rd a_j}dS_{\SW}\right|_{z=\const} 
    = d\omega_j. 
\eeqn
The prepotential $\calF = \calF(a_1,\cdots,a_N)$ is 
defined by the differential equations 
\beqn
    \frac{\rd \calF}{\rd a_j} = a_j^D. 
\eeqn
The matrix elements $\calP_{jk}$ of $\calP$ can be 
expressed as second derivatives of the prepotential:
\beqn
    \calP_{jk} = \frac{\rd^2 \calF}{\rd a_j \rd a_k}. 
\eeqn

\subsection{Quotient curve of genus $N$}

The Prym variety $\Prym(C/C_2)$ is isogenous to the Jacobi 
variety $\Jac(C')$ of the quotient curve $C' = C/\sigma'$ 
by the following involution $\sigma'$: 
\begin{itemize}
    \item Case A: $\sigma' = \sigma_2$. 
    \item Case B: $\sigma' = \sigma_1 \sigma_2$. 
\end{itemize}
The quotient curve $C'$ is also hyperelliptic and has 
genus $N$.  The matrix $\calP$ is actually the period matrix 
of the Jacobi variety $\Jac(C')$: 
\beqn
    \Jac(C') \simeq \bbC^N / (\bbZ^N + \calP \bbZ^N). 
\eeqn
The differentials $dS_{\SW}$ and $d\omega_j$ are ``even'' 
(i.e., invariant) under the action of $\sigma'$, so that 
they are the pull-back of differentials on $C'$.  On the 
other hand, the homology classses $[A_j]$ and $[B_j]$
are mapped by the projection $p':C \to C'$ to an integer 
multiple of the homology classes $[A'_j]$ and $[B'_j]$ of 
a symplectic basis on $C'$.  More precisely, 
\beqn
    p'([A_j]) = [A'_j], \quad 
    p'([B_j]) = \frac{2}{d_j}[B'_j]. 
\eeqn
This is the origin of the factor $d_j/2$ in the period 
integral representation of the matrix elements of $\calP$.  

The equation of the quotient curve $C'$ can be written 
out in terms of the following two invariants 
$\xi$ and $\eta$ of $\sigma'$: 
\begin{itemize}
    \item Case A: $\xi = x^2, \  \eta = y$. 
    \item Case B: $\xi = x^2, \  \eta = xy$. 
\end{itemize}
The following equations of the quotient curves $C'$ 
are thus derived. 
\begin{enumerate}
\item For Seiberg-Witten curves: 
\beqn
    SO(2N+1) &:& 
      \eta^2 = \xi \Bigl( Q(\xi^2) - 4 \mu^2 \xi \Bigr). 
      \nonumber \\
    Sp(2N) &:& 
      \eta^2 = Q(\xi) \Bigl( \xi Q(\xi) + 4 \mu \Bigr). 
      \nonumber \\
    SO(2N) &:& 
      \eta^2 = \xi \Bigl( Q(\xi)^2 - 4 \mu^2 \xi^2 \Bigr). 
    \nonumber 
\eeqn
\item For the other Toda curves: 
\beqn
    B_N^{(1)},C_N^{(1)} &:& 
      \eta^2 = \xi \Bigl( Q(\xi)^2 - 4 \mu^2 \Bigr). 
      \nonumber \\
    A_{2N}^{(2)} &:& 
      \eta^2 = \xi Q(\xi)^2 - 4 \mu^2. 
      \nonumber 
\eeqn
\end{enumerate}

Note that the curves $C$ and $C'$ exhibit somewhat different 
characteristics.  The curve $C$, viewed as a double covering 
of the $x$-sphere, has two points $P_\infty$ and $Q_\infty$ 
at infinity above $x = \infty$.  These two points correspond 
to $z = \infty$ and $z = 0$, and mapped to each other by 
the hyperelliptic involution $\sigma_1$.  This is a typical 
characteristic of the spectral curves of affine Toda systems 
\cite{bib:Kr-Toda,bib:vM-Mu,bib:Ad-vM}. 
The curve $C'$, in contrast, has a single point at infinity 
above the $\xi$-shere.  In particular, $C'$ is branched 
over $\xi = \infty$.  Hyperelliptic curves of this type 
arise in the KdV  hierarchy 
\cite{bib:Du-Ma-No-KdV,bib:Kr-KdV}.  As well known, 
the KdV hierarchy is a special case of the KP hierarchy 
with only the ``odd'' time variables $t_{2n+1}$ being left 
nontrivial \cite{bib:Se-Wi-KP}.  

The $SU(2)$ curve is however exceptional: It has the 
involution $\sigma_2$.  By this accidental symmetry, 
one can construct the quotient curve $C' = C/\sigma'$ 
with $\sigma' = \sigma_1\sigma_2$, which can be written 
\beqn
    \eta^2 = \xi \Bigl( (\xi - u_2)^2 - 4 \mu^2 \Bigr) 
\eeqn
in terms of the invariants $\xi = x^2$ and $\eta = xy$. 
By shifting $\xi \mapsto \xi - u_2$, this turns into 
the substantially the same curve 
\beqn
    \eta^2 = (\xi + u_2) (\xi - 2 \mu) (\xi + 2 \mu) 
\eeqn
as Seiberg and Witten first derived \cite{bib:Se-Wi}.  
As Gorsky et al. noted \cite{bib:GKMMM}, this curve 
appears in a classical study on modulations of 
elliptic solutions of the KdV equation.

%%%%%%%%%%%%%%%%%%%%%%%%%%%%%%%%%%%%%%%%%%%%%%%%%%%%%%%%%%%%%%%%%%%%

\section{Whitham Deformations of $SU(N+1)$ curve}
\setcounter{equation}{0}

\subsection{Setup and results of Gorsky et al.}

The Whitham deformations for the $SU(N+1)$ Seiberg-Witten curve 
takes the form 
\beqn
    \left.\frac{\rd}{\rd a_j}dS\right|_{z=\const} = d\omega_j, 
    \quad 
    \left.\frac{\rd}{\rd T_n}dS\right|_{z=\const} = d\Omega_n. 
    \label{eq:Whitham}
\eeqn
The precise setup is as follows: 
\begin{enumerate}
\item 
The moduli $\vec{u} = (u_1,\cdots,u_N)$ are now 
understood to be functions $u_j(\vec{a},\vec{T})$ 
of $\vec{a} = (a_1,\cdots,a_N)$ and 
$\vec{T} = (T_1,T_2,\cdots)$.  The new parameters 
$T_n$ are the ``slow variables'' in the theory of 
Whitham equations.  These deformed moduli 
$u_j = u_j(\vec{a},\vec{T})$ are required to reduce 
to the Seiberg-Witten moduli $u_j = u_j(\vec{a})$ 
(i.e., the inverse of the period map $\vec{u} \to \vec{a}$ 
from the $\vec{u}$-space to the $\vec{a}$-space) at 
$\vec{T} = (1,0,0,\cdots)$.  
\item 
$d\Omega_n$ are meromorphic differentials of the 
second kind with poles at the two points $P_\infty$ 
and $Q_\infty$ only, and normalized to have zero 
$A_j$ periods: 
\beqn
    \oint_{A_j} d\Omega_n = 0 \quad (j = 1,\cdots,N). 
\eeqn
\item 
$d\omega_j$ are the normalized holomorphic differentials 
on the curve $C$. 
\item 
The differential $dS$ is a linear combination of 
these differentials of the form 
\beqn
    dS = \sum_{n\ge 1} T_n d\Omega_n 
       + \sum_{j=1}^N a_j d\omega_j, 
\eeqn
and required to reduce to the Seiberg-Witten differential 
$dS_{\SW}$  at the point $\vec{T} = (1,0,0,\cdots)$. 
\end{enumerate}

The solution of Gorsky et al. \cite{bib:GMMM} for the 
$SU(N+1)$ Seiberg-Witten curve is constructed by the 
following steps. 
\begin{enumerate} 
\item 
Consider the meromorphic differentials 
\beqn
    d\Omegahat_n = R_n(x) \frac{dz}{z}, \quad 
    R_n(x) = \Bigl( P(x)^{n/(N+1)} \Bigr)_{+}. 
\eeqn
Here $(\cdots)_{+}$ denotes the polynomial part of 
a Laurent series of $x$.  The fractional power of $P(x)$ 
is understood to be a Laurent series of the form 
$x^n + \cdots$ at $x = \infty$. Since $R_1(x) = x$, 
$d\Omegahat_1$ is nothing but the Seiberg-Witten 
differential.  As in the case of the Seiberg-Witten 
differential (\ref{eq:dSSW-by-u}), the $u$-derivatives 
of these meromorphic differentials turn out to be 
holomorphic differentials. 
\item 
Consider the differential 
\beqn
    dS = \sum_{n\ge 1} T_n d\Omegahat_n 
\eeqn
and its period integrals 
\beqn
    a_j = \oint_{A_j} dS 
    = \sum_{n\ge 1} T_n \oint_{A_j}d\Omegahat_n. 
\eeqn
These period integrals are functions of the moduli 
$u_j$ and the deformation parameters $T_n$.  
They determine a family of deformations of the 
Seiberg-Witten period map $\vec{u} \mapsto \vec{a}$ 
with parametes $T_n$.  
\item 
The period map $\vec{u} \mapsto \vec{a}$ from the 
$\vec{u}$-space to the $\vec{a}$-space is invertible 
if $\vec{T}$ is close to $(1,0,0,\cdots)$, 
because the Seiberg-Witten period map at this point 
is invertible.  The inverse map $\vec{a} \mapsto \vec{u} = 
\Bigl(u_1(\vec{a},\vec{T}),\cdots,u_N(\vec{a},\vec{T})\Bigr)$ 
gives deformations of the Seiberg-Witten moduli 
$u_j = u_j(\vec{a})$, hence of the curve $C$, with 
parameters $T_n$.  
\item 
The differentials 
\beqn
    d\Omega_n = d\Omegahat_n 
      - \sum_{j=1}^N c_j^{(n)} d\omega_j, \quad 
    c_j^{(n)} = \oint_{A_j} d\Omegahat_n 
\eeqn
satisfy the required normalization condition. 
\item  
$dS$ is now a linear combination of $d\Omega_n$ 
and $d\omega_j$ of the required form. 
\end{enumerate}
The outcome is the following theorem: 

\begin{theorem}
The differential $dS$ satisfies the Whitham equations 
(\ref{eq:Whitham}) under the deformations of the curve 
$C$ thus constructed. 
\end{theorem}

We review the proof of this result in the rest of 
this section.

\subsection{Differentiating $d\Omegahat_n$ by moduli} 

The first, and most essential step is to derive the 
following property of $d\Omegahat_n$.  Note that $u_j$'s 
and $T_n$'s are now understood to be independent variables.  

\begin{lemma}
$(\rd/\rd u_j)d\Omegahat_n |_{z=\const}$ 
are holomorphic differentials on $C$. 
\end{lemma}

\proof
We first derive a set of conditions that the polynomial 
$R_n(x)$ should satisfy, and verify that they are indeed 
fulfilled.  Differentiating the eqution of the curve $C$ 
gives the relation 
\beqn
    \frac{\rd P(x)}{\rd u_j} 
    + \frac{\rd P(x)}{\rd x} 
      \left.\frac{\rd x}{\rd u_j}\right|_{z=\const} 
    = 0. 
\eeqn
Therefore, recalling that $dz/z = P'(x)dx/y$, one can rewrite 
the $u_j$-derivative of $d\Omegahat_n$ as follows: 
\beqn
    \left.\frac{\rd}{\rd u_j}d\Omegahat_n\right|_{z=\const} 
    &=& \left( \frac{\rd R_n(x)}{\rd u_j} 
        + R_n'(x)\left.\frac{\rd x}{\rd u_j}\right|_{z=\const} 
        \right) \frac{dz}{z}
    \nonumber \\
    &=& \left( \frac{\rd R_n(x)}{\rd u_j} 
        + R_n'(x) \frac{x^{N+1-j}}{P'(x)} 
        \right) \frac{P'(x)dx}{y} 
    \nonumber \\
    &=& \left( \frac{\rd R_n(x)}{\rd u_j} P'(x) 
        + R_n'(x) x^{N+1-j} 
        \right) \frac{dx}{y}. 
\eeqn
For this differential to be a holomorphic differential on $C$, 
therefore, the polynomial $R_n(x)$ has to satisfy the following 
condition:
\beqn
    \deg \left( \frac{\rd R_n(x)}{\rd u_j} P'(x) 
    + R_n'(x) x^{N+1-j} \right) \le N - 1. 
\eeqn
Let us confirm that $R_n(x) = \Bigl( P(x)^{n/(N+1)} \Bigr)_{+}$ 
has this property.  By the definition of $(\cdots)_{+}$, 
$R_n(x)$ can be written 
\beqn
    R_n(x) = P(x)^{n/(N+1)} + O(x^{-1}). 
    \nonumber 
\eeqn
Differentiating this relation by $x$ and $u_j$, respectively, 
gives 
\beqn
    R_n'(x) &=& 
      \frac{n}{N+1} P(x)^{(n-N-1)/(N+1)} P'(x) + O(x^{-2}), 
    \nonumber \\
    \frac{\rd R_n(x)}{\rd u_j} &=& 
      - \frac{n}{N+1} P(x)^{(n-N-1)/(N+1)} x^{N+1-j} + O(x^{-1}). 
\eeqn
From these relations, one can easily see that the above 
condition is certainly satisfied.  \qed

\subsection{Deriving Whitham equations} 

Once $d\Omegahat_n$ turns out to have the aforementioned 
property, deriving the Whitham equations (\ref{eq:Whitham}) 
is rather straightforward.  Let us present this calculation 
following the work of Itoyama and Morozov \cite{bib:It-Mo}. 

First, since  that $d\omega_j$ $(j = 1,\cdots,N$) give 
a basis of the space of holomorphic differentials, 
the holomorphic differentials in the above Lemma 
can be written 
\beqn
    \frac{\rd}{\rd u_k} 
     \left.d\Omegahat_n \right|_{z=\const}
    = \sum_{j=1}^N \sigma^{(n)}_{kj} d\omega_j. 
    \label{eq:dOm-by-u}
\eeqn
The coefficients are determined by integrating 
the both hand sides along $A_j$: 
\beqn
    \sigma^{(n)}_{kj} 
    = \oint_{A_j} \frac{\rd}{\rd u_k} 
        \left.d\Omegahat_n \right|_{z=\const} 
    = \frac{\rd}{\rd u_k} \oint_{A_j} d\Omegahat_n 
    = \frac{\rd c^{(n)}_j}{\rd u_k}. 
    \label{eq:sigma-nkj}
\eeqn

Second, the derivatives of $a_j = a_j(\vec{u},\vec{T})$ 
turn out to be written 
\beqn
    \frac{\rd a_j}{\rd u_k} = \sum_{n\ge 1} T_n \sigma^{(n)}_{kj}, 
    \quad 
    \frac{\rd a_j}{\rd T_n} = c^{(n)}_j. 
    \label{eq:a-by-uT}
\eeqn
This can be readily verified by directly differentiating 
\beqn
    a_j = \oint_{A_j} dS = \sum_{n\ge 1} T_n c^{(n)}_j 
\eeqn
and using (\ref{eq:sigma-nkj}). 

Now change the independent variables from 
$(\vec{u},\vec{T})$ to $(\vec{a},\vec{T})$. 
These two systems of coordinates are connected by 
the functions $a_j = a_j(\vec{u},\vec{T})$ and 
$u_k = u_k(\vec{a},\vec{T})$. In this setup, 
the following identities are satisfied: 
\beqn
    \sum_{k=1}^N \frac{\rd u_k}{\rd a_i} \frac{\rd a_j}{\rd u_k} 
      = \delta_{ij}, 
    \quad 
    \sum_{k=1}^N \frac{\rd u_k}{\rd T_m} \frac{\rd a_j}{\rd u_k} 
      = - \frac{\rd a_j}{\rd T_m}. 
    \label{eq:uaT-chain}
\eeqn
The first relation is obvious from the chain rule.  
The second one is rather confusing; this is obtained 
by differentiating the identity 
\beqn
    a_j = a_j(u_1(\vec{a},\vec{T}), \cdots, 
              u_N(\vec{a},\vec{T}), \vec{T}) 
    \nonumber 
\eeqn
by $T_m$. 

Using the chain rule along with these relations  
(\ref{eq:dOm-by-u}), (\ref{eq:a-by-uT}) and 
(\ref{eq:uaT-chain}), one can verify the Whitham 
equations (\ref{eq:Whitham}) as follows.
\begin{enumerate}
\item The first equation of (\ref{eq:Whitham}): 
\beqn
    \left.\frac{\rd}{\rd a_i}dS \right|_{z=\const} 
    &=& \sum_{n} T_n \left. \frac{\rd}{\rd a_i} 
        d\Omegahat_n \right|_{z=\const} 
    \nonumber \\
    &=& \sum_{n,k} T_n \frac{\rd u_k}{\rd a_i} 
        \left. \frac{\rd}{\rd u_k} 
        d\Omegahat_n \right|_{z=\const} 
    \nonumber \\
    &=& \sum_{j,k} \frac{\rd u_k}{\rd a_i} 
        \cdot \sum_{n} T_n \sigma^{(n)}_{kj} 
        \cdot d\omega_j 
    \nonumber \\
    &=& \sum_{j,k} \frac{\rd u_k}{\rd a_i} 
        \frac{\rd a_j}{\rd u_k} d\omega_j
    \nonumber \\
    &=& d\omega_i. 
    \nonumber 
\eeqn
\item The second equation of (\ref{eq:Whitham}): 
\beqn
    \left.\frac{\rd}{\rd T_m}dS \right|_{z=\const} 
    &=& d\Omegahat_m + 
        \sum_{n} T_n \left. \frac{\rd}{\rd T_m} 
        d\Omegahat_n \right|_{z=\const} 
    \nonumber \\
    &=& d\Omegahat_m + 
        \sum_{n,k} T_n \frac{\rd u_k}{\rd T_m} 
        \left. \frac{\rd}{\rd u_k} 
        d\Omegahat_n \right|_{z=\const}
    \nonumber \\
    &=& d\Omegahat_m + 
        \sum_{j,k} \frac{\rd u_k}{\rd T_m} 
        \cdot \sum_{n} T_n \sigma^{(n)}_{kj} 
        \cdot d\omega_j 
    \nonumber \\
    &=& d\Omegahat_m + 
        \sum_{j,k} \frac{\rd u_k}{\rd T_m} 
        \frac{\rd a_j}{\rd u_k} d\omega_j 
    \nonumber \\
    &=& d\Omegahat_m - \sum_{j} c^{(m)}_j d\omega_j 
    \nonumber \\
    &=& d\Omega_m. 
    \nonumber 
\eeqn
\end{enumerate}
This completes the proof of the Theorem.

\subsection{Prepotential and homogeneity} 

The prepotential $\calF = \calF(a,T)$ is defined by the 
equations 
\beqn
    \frac{\rd \calF}{\rd a_j} = \oint_{B_j} dS, \quad 
    \frac{\rd \calF}{\rd T_n} = 
      - \oint_{P_\infty} f_n(z) dS 
      - \oint_{Q_\infty} g_n(z) dS. 
    \label{eq:define-F}
\eeqn
$P_\infty$ and $Q_\infty$ are the two points at infinity 
($z = \infty$ and $z = 0$); $\oint_{P_\infty}$ and 
$\oint_{Q_\infty}$ are integrals along a small closed 
path encircling the indicated point once in the 
anti-clockwise direction.  $f_n(z)$ and $g_n(z)$ are 
Laurent polynomials that represent the singular part of 
$d\Omega_n$ at $P_\infty$ and $Q_\infty$: 
\beqn
    d\Omega_n &=& df_n(z) + \mbox{holomorphic} \quad 
      (P \to P_\infty), 
    \nonumber \\
    d\Omega_n &=& dg_n(z) + \mbox{holomorphic} \quad 
      (P \to Q_\infty). 
\eeqn
These Laurent polynomials have constant coefficients for 
all $n$. \footnote{Our previous paper \cite{bib:Ta-u-plane} 
contains a wrong comment on this fact.}  To see this, 
let us note that the fractional power of $P(x)$ in 
$R_n(x)$ can be written 
\beqn
    P(x)^{n/(N+1)} = \Bigl(z + \frac{\mu^2}{z}\Bigr)^{n/(N+1)}. 
\eeqn
The singular part of Laurent expansion of the right 
hand side at $z = \infty$ or $z = 0$ determines 
the Laurent polynomials $f_n(z)$ and $g_n(z)$. 
Obviously the singular part is a Laurent polynomial 
with constant coefficients, and accordingly $f_n(z)$ 
and $g_n(z)$, too, turn out to have constant coefficients. 
The compatibility of the above defining equations for 
$\calF$ is ensured by Riemann's bilinear relations.  
Second derivatives of the prepotential are also related 
to period integrals: 
\beqn
    && 
    \frac{\rd^2 \calF}{\rd a_j \rd a_k} = 
      \oint_{B_j} d\omega_k, 
    \quad 
    \frac{\rd^2 \calF}{\rd a_j \rd T_n} = 
      \oint_{B_j} d\Omega_n, 
    \nonumber \\
    && 
    \frac{\rd^2 \calF}{\rd T_m \rd T_n} =  
      - \oint_{P_\infty} f_m(z) d\Omega_n 
      - \oint_{Q_\infty} g_m(z) d\Omega_n. 
\eeqn

The construction of the Whitham deformations also ensures the 
homogeneity 
\beqn
    \sum_{n\ge 1} T_n \frac{\rd \calF}{\rd T_n} 
    + \sum_{j=1}^N a_j \frac{\rd \calF}{\rd a_j} 
    = 2 \calF. 
\eeqn
To see this, first note that the period integrals 
$a_j = a_j(\vec{u},\vec{T})$ have the obvious homogeneity 
\beqn
    a_j(\vec{u},\lambda\vec{T}) = \lambda a_j(\vec{u},\vec{T}). 
\eeqn
Acoordingly, $u_j = u_j(\vec{a},\vec{T})$ are invariant under 
the rescaling of $a_j$ and $T_n$, i.e., they are 
homogeneous functions of degree zero: 
\beqn
    u_j(\lambda \vec{u},\lambda \vec{T}) = u_j(\vec{u},\vec{T}). 
\eeqn
One can see, from these fact, that the period integrals 
on the right hand side of (\ref{eq:define-F}) are 
homogeneous functions of degree one, so that the prepotential 
becomes homogeneous of degree two (upon suitably normalizing 
the linear part).

%%%%%%%%%%%%%%%%%%%%%%%%%%%%%%%%%%%%%%%%%%%%%%%%%%%%%%%%%%%%%%%%%%%%

\section{Whitham Deformations of $SO(2N)$ curve}
\setcounter{equation}{0}

\subsection{Setup and results}

As warm-up, let us examine some fundamental properties 
of the Seiberg-Witten differential $dS_{\SW}$. 
The $u_j$-derivatives of $dS_{\SW}$ can be written 
\beqn
    \left.\frac{\rd}{\rd u_j} dS_{\SW} \right|_{z=\const} 
    &=& \frac{x^{2N-2j-2}dx}{\sqrt{x^{-4}Q(x^2)^2 - 4\mu^2}} 
    \nonumber \\
    &=& \frac{x^{2N-2j}dx}{y} 
        \quad (j = 1,\cdots,N). 
\eeqn
As expected, they are holomorphic differentials on $C$. 
Furthermore, like $dS_{\SW}$ itself, they are invariant 
under the involution $\sigma':(x,y) \mapsto (-x,-y)$, 
and can be identified with holomorphic differentials 
on the quotient curve $C' = C/\sigma'$ with coordinates 
$\xi = x^2$ and $\eta = xy$:
\beqn
    \left.\frac{\rd}{\rd u_j} dS_{\SW} \right|_{z=\const} 
    = \frac{\xi^{N-j}d\xi}{2\eta} 
    \quad (j=1,\cdots,N). 
\eeqn

Following the method of Gorsky et al., we now seek for 
a series of meromorphic differentials $d\Omegahat_n$ 
of the form 
\beqn
    d\Omegahat_n = R_n(x) \frac{dz}{z}, \quad 
    R_n(x) = \mbox{polynomial}, 
\eeqn
with the same properties.  As one can see by careful 
inspection of the proof in the $SU(N+1)$ Seiberg-Witten 
curve, the fractional power construction persists to be 
meaningful even if the polynomial $P(x)$ is replaced by 
a rational (or Laurent series)  superpotential $W(x)$. 
A new feature in the present setup is the parity; 
for the differential to be invariant under $\sigma'$, 
the polynomial $R_n(x)$ has to be odd, 
\beqn
    R_n(-x) = - R_n(x). 
\eeqn
The superpotential $W(x)$ is even, and the leading 
term is $x^{2N-2}$. We are thus led to the following: 
\beqn
    R_n(x) = \Bigl( W(x)^{(2n-1)/(2N-2)} \Bigr)_{+} 
    \quad (n = 1,2,\cdots). 
\eeqn
Note that these polynomials are indeed odd polynomials. 
Furthermore, $R_1(x) = x$, so that in this case, too, 
$d\Omegahat_1$ is equal to the Seiberg-Witten differential. 

\begin{lemma}
$(\rd/\rd u_j) d\Omegahat_n |_{z=\const}$ 
are holomorphic differentials on $C$, and 
invariant under the involution $\sigma'$. 
\end{lemma}

\proof
We can proceed in mostly the same way as the case of the 
$SU(N+1)$ Seiberg-Witten curve.  The $u_j$-derivatives of 
$d\Omegahat_n$ can be written 
\beqn
    \left.\frac{\rd}{\rd u_j} d\Omegahat_n\right|_{z=\const} 
    = \left( x^2 \frac{\rd R_n(x)}{\rd u_j} W'(x) 
        - x^2 R_n'(x) \frac{\rd W(x)}{\rd u_j} \right)
        \frac{dx}{y}. 
\eeqn
The proof now boilds down to verifying the following two 
statements: 
\begin{enumerate}
\item The factor in front of $dx/y$ is a polynomial. 
\item The degree of this polynomial does not exceed $2N - 2$. 
\end{enumerate}
Let us consider the first statement.  The problem is that 
$\rd W(x)/\rd u_j$ and $W'(x)$ respectively have a pole of 
second and third order at $x = 0$.  $\rd R_n(x)/\rd u_j$ and 
$R_n'(x)$, however, have a zero at $x = 0$, because $R_n(x)$ 
is an odd polynomial and has no constant term.  Therefore, 
along with the factor $x^2$, they cancel the pole of at most 
third order of the other two functions.  The second statement 
can be verified by the same reasoning as the proof for the 
the $SU(N+1)$ Seiberg-Witten curve.  \qed

\subsection{Whitham deformations}

Having obtained the meromorphic differentials 
$d\Omegahat_n$ with the aforementioned properties, 
we can now construct a family of Whitham deformations. 

The construction is parallel to the case of 
the $SU(N+1)$ Seiberg-Witten curve: 
\begin{enumerate}
\item  
Consider the differential 
\beqn
    dS = \sum_{n\ge 1} T_n d\Omegahat_n. 
\eeqn
and its period integrals 
\beqn
    a_j = \oint_{A_j} dS 
    = \sum_{n\ge 1} T_n \oint_{A_j} d\Omegahat_n. 
\eeqn
The period integrals $a_j = a_j(\vec{u},\vec{T})$ 
determine a period map $\vec{u} \mapsto \vec{a}$ 
from the $\vec{u}$-space to the $\vec{a}$-space. 
\item 
The period map $\vec{u} \mapsto \vec{a}$ is invertible 
if $\vec{T}$ is close to $(1,0,0,\cdots)$.  
The inverse map $\vec{a} \mapsto \vec{u} = 
\Bigl(u_1(\vec{a},\vec{T}),\cdots,u_N(\vec{a},\vec{T})\Bigr)$ 
gives deformations of the curve $C$ with parameters $\vec{T}$. 
\item 
The differentials 
\beqn
    d\Omega_n = d\Omegahat_n - \sum_{j=1}^N c_j^{(n)} d\omega_j, 
    \quad 
    c_j^{(n)} = \oint_{A_j} d\Omegahat_n, 
\eeqn
satisfy the normalization condition 
\beqn
    \oint_{A_j} d\Omega_n = 0 \quad (j = 1,\cdots,N).  
\eeqn
\item 
Eventually, the differential $dS$ can be written 
\beqn
    dS = \sum_{n\ge 1} T_n d\Omega_n + \sum_{j=1}^N a_j d\omega_j. 
\eeqn
\end{enumerate}
The following can be proven in exactly the same way as
the proof for the $SU(N+1)$ Seiberg-Witten curve: 

\begin{theorem}
The differential $dS$ satisfies the Whitham equations 
\beqn
    \left.\frac{\rd}{\rd a_j} dS\right|_{z=\const} = d\omega_j, 
    \quad 
    \left.\frac{\rd}{\rd T_n} dS\right|_{z=\const} = d\Omega_n, 
\eeqn
under the deformations of the curve $C$ thus constructed. 
\end{theorem}

\subsection{Relation to KdV hierarchy} 

By construction, the differentials $dS$, $d\Omegahat_n$, 
$d\Omega_n$ and $d\omega_j$ are all invariant under the 
involution $\sigma'$. Accordingly, they actually descend to 
(or, equivalently, are the pull-back of) differentials on 
the quotient curve $C' = C/\sigma'$. In particular, the 
counterpart of $d\omega_j$ are also holomorphic, and, as 
already mentioned, form a normalized basis of holomorphic 
differentials on $C'$. 

The meromorphic differentials $d\Omega_n$ possess an 
even more interesting interpretation.  Thy correspond 
to meromorphic differentials on $C'$ with a single pole 
at $\xi = \infty$ (which is the image of $P_\infty$ and 
$Q_\infty$).  Recall that this is a branch point of the 
covering map of $C'$ over the $\xi$-sphere. This is 
substantially the same setup that emerges in hyperelliptic 
solutions of the KdV hierarchy 
\cite{bib:Du-Ma-No-KdV,bib:Kr-KdV}.  Those hyperelliptic 
solutions are constructed from a theta function and 
a series of meromorphic differentials $d\Omega^{\KdV}_n$ 
($n = 1,2,\cdots$) with a single pole (of order $2n$) 
at a fixed branch point (such as the point $\xi = \infty$) 
of the hyperelliptic curve. 

In this respect, we should have numbered the deformation 
variables as $T_1,T_3,T_5,\cdots$ rather than 
$T_1,T_2,T_3,\cdots$, following the usual numbering 
of the time variables in the KdV hierarchy.  Note that 
the odd indices correspond to the degrees $2n-1$ of 
powers of $W(x)^{1/(2N-2)}$ in the definition of $R_n(x)$. 

We must, however, also add that our meromorphic 
differentials are not exactly the same as those 
in the standard formulation of the KdV hierarchy.  
The meromorphic differentials $d\Omega^{\KdV}_n$ 
in the standard setup are normalized as 
\beqn
    d\Omega^{\KdV}_n = d\xi^{n-(1/2)}  
      + \mbox{holomorphic} 
\eeqn
at $\xi = \infty$.  Our meromorphic differentials 
$d\Omega_n$ are not of this form; they are a linear 
combination of $d\Omega^{\KdV}_n$.  Accordingly, 
the ``fast'' and ``slow'' time variables are also 
a linear combination of the standard ones.  

These remarks also apply to the other cases where 
the essential part of the theory is described by 
the quotient curve $C'$ of the KdV type.

%%%%%%%%%%%%%%%%%%%%%%%%%%%%%%%%%%%%%%%%%%%%%%%%%%%%%%%%%%%%%%%%%%%%

\section{Non-simply-Laced Gauge Groups and Some Other Cases}
\setcounter{equation}{0}

\subsection{$SO(2N+1)$ and $Sp(2N)$}

The case of the non-simply-laced gauge groups 
$SO(2N+1)$ and $Sp(2N)$ can be treated in almost 
the same way as the case of $SO(2N)$.  The curve $C$ 
has the involution $\sigma'$, and the quotient curve 
$C' = C/\sigma'$ has genus $N$. The Seiberg-Witten 
differential $dS_{\SW}$ is invariant under this 
involution.  Our first task is to construct 
meromorphic differentials 
\beqn
    d\Omegahat_n = R_n(x) \frac{dz}{z}, \quad 
    R_n(x) = \mbox{odd polynomial}, 
\eeqn
whose $u_j$-derivatives are holomorphic differentials 
on $C$.  

The polynomials $R_n(x)$ are again given by the polynomial 
part of fractional powers of the superpotential $W(x)$: 
\beqn
    SO(2N+1) &:& 
      R_n(x) = \Bigl(W(x)^{(2n-1)/(2N-1)}\Bigr)_{+}. 
    \\
    Sp(2N)   &:&  
      R_n(x) = \Bigl(W(x)^{(2n-1)/(2N+2)}\Bigr)_{+}. 
\eeqn 
The $u_j$-derivatives can be written as follows: 
\begin{enumerate}
\item For $SO(2N+1)$, 
\beqn
    \frac{\rd}{\rd u_j} \left.d\Omegahat_n\right|_{z=\const} 
    &=& \left( x \frac{\rd R_n(x)}{\rd u_j} W'(x) 
        - x R_n'(x) \frac{\rd W(x)}{\rd u_j} \right)
        \frac{dx}{y}. 
\eeqn
\item For $Sp(2N)$, 
\beqn
    \frac{\rd}{\rd u_j} \left.d\Omegahat_n\right|_{z=\const} 
    &=& \left( x^{-1} \frac{\rd R_n(x)}{\rd u_j} W'(x) 
        - x^{-1} R_n'(x) \frac{\rd W(x)}{\rd u_j} \right)
        \frac{dx}{y}. 
\eeqn
\end{enumerate}
The prefactor of $dx/y$ turns out to be a polynomial 
of degree $\le 2N-2$ and $\le 2N-1$ for $SO(2N+1)$ and 
$Sp(2N)$, respectively. Therefore the above differentials 
are holomorphic differentials on $C$.

The Whitham deformations of $C$ are given by the inverse 
period map $\vec{a} \mapsto \vec{u}$ of the period integrals 
\beqn
    a_j = \oint_{A_j} dS 
    = \sum_{n\ge 1} T_n \oint_{A_j} d\Omegahat_n. 
\eeqn
of the differential
\beqn
    dS = \sum_{n\ge 1} T_n d\Omegahat_n. 
\eeqn
The normalized meromorphic differentials $d\Omega_n$ are 
given by 
\beqn
    d\Omega_n = d\Omegahat_n - \sum_{j=1}^n c_j^{(n)} d\omega_j, 
    \quad 
    c_j^{(n)} = \oint_{A_j} d\Omegahat_n. 
\eeqn
Now $dS$ can be written 
\beqn
    dS = \sum_{n\ge 1} T_n d\Omega_n + \sum_{j=1}^N a_j d\omega_j, 
\eeqn
and satisfy the Whitham equations 
\beqn
    \left.\frac{\rd}{\rd a_j} dS\right|_{z=\const} = d\omega_j, 
    \quad 
    \left.\frac{\rd}{\rd T_n} dS\right|_{z=\const} = d\Omega_n 
\eeqn
under the deformations of the curve $C$ by the inverse 
period map $\vec{a} \mapsto \vec{u}$.

\subsection{Some other cases}

The same fractional power construction of Whitham deformations 
also applies to the Toda curves for other classical affine Lie 
algebras, namely, $B_N^{(1)}$, $C_N^{(1)}$ and $A_{2N}^{(2)}$.  
The meromorphic differentials $d\Omegahat_n$ are obtained in 
the same form as the other cases. The polynomials $R_n(x)$ 
are given by the following:
\beqn
    B_N^{(1)}, C_N^{(1)} &:& 
      W(x) = \Bigl( W(x)^{(2n-1)/(2N)} \Bigr)_{+}, 
    \nonumber \\
    A_{2N}^{(2)} &:& 
      W(x) = \Bigl( W(x)^{(2n-1)/(2N+1)} \Bigr)_{+}. 
\eeqn

The case of the $A_{2N}^{(2)}$ (also called $BC_N$) 
Toda curve, seems to be particularly interesting and 
mysterious. Although this does not appear in the list 
of ordinary $\calN = 2$ SUSY Yang-Mills theories, an 
M-theoretical interpretation \cite{bib:Wi-M5} might 
be possible.  

Actually, the same construction works with no problem for 
a more general rational superpotential, e.g., 
\beqn
    W(x) = P(x) + \sum_{k=1}^M \frac{v_k}{x - c_k}. 
\eeqn
Note that the curve $C$ is still hyperelliptic.  
This is a special case of M5-branes with semi-infinite 
D4 branes (arising from the poles of $W(x)$) on both sides 
of the stack of two NS 5-branes.

\subsection{Yang-Mills theories with fundamental matters}

The Seiberg-Witten curve does not take the previous 
form (\ref{eq:SW-curve}) if fundamental matters 
(hypermultiplets in the fundamental representation) 
exist.  One can, however, rewrite the curve in this form 
if $W(x)$ can be an irrational function.  Let us examine 
this prescription in the case of the $SU(N+1)$ Yang-Mills 
theory with fundamental matters.  

If the theory contains $N_f$ ($<2N$) fundamental matters, 
the Seiberg-Witten curve takes the form 
\beqn
    z + \frac{\mu^2 R(x)}{z} = P(x), 
\eeqn
where $R(x)$ is the polynomial 
\beqn
    R(x) = \prod_{i=1}^{N_f} (x + m_i). 
\eeqn
This curve can be formally converted into the 
form of (\ref{eq:SW-curve}) by changing coordinates 
from $x$ and $z$ to $x$ and 
\beqn
    w = z / \sqrt{R(x)}. 
\eeqn
The converted curve can be written 
\beqn
    w + \frac{\mu^2}{w} = W(x) = \frac{P(x)}{\sqrt{R(x)}}, 
\eeqn
thus an irrational superpotential arises.  
The role of $z$ is now played by $w$. For instance, 
the Seiberg-Witten differential can be written 
\beqn
    dS_{\SW} = x\frac{dw}{w}.
\eeqn

Since the superpotential itself is multi-valued, 
The fractional power construction requires a more 
careful handling.  This multi-valuedness is however 
relatively harmless, simply affecting an overall 
multiplicative constant of the meromorphic 
differentials $d\Omegahat_n$.

%%%%%%%%%%%%%%%%%%%%%%%%%%%%%%%%%%%%%%%%%%%%%%%%%%%%%%%%%%%%%%%%%%%%

\section{Topologically Twisted Gauge Theories}
\setcounter{equation}{0}

\subsection{Donaldson-Witten function and $u$-plane integral}

Correlation functions of topologically twisted gauge 
theories on a four-dimensional manifold $X$ can be 
collected into a generating function.  This generating 
function is called the Donaldson-Witten function.  
For instance, consider a 2-cycle observable $I(S)$, 
$S \in H_2(X,\bbZ)$, and a 0-cycle observable $\calO(P)$, 
$P \in H_0(X,\bbZ)$, of the form 
\beqn
    I(S) = \const \int_S G^2 \Tr \phi^2, \quad 
    \calO(P) = \sum_k c_k \Tr \phi^k(P). 
\eeqn
$G$ is a transformation that generates a standard 
solution of the descent equations for observables 
\cite{bib:Mo-Wi,bib:Ma-Mo}.  The Donaldson-Witten 
function of these observables is the path integral 
\beqn
     Z_{\DW} = \Bigl< \exp\Bigl(I(S) + \calO(P)\Bigr)\Bigr>. 
\eeqn

In the case with $b_2^{+}(X) = 1$, the Donaldson-Witten 
function becomes a  sum of two pieces: 
\beqn
    Z_{\DW} = Z_{\SW} + Z_u. 
\eeqn
The first piece $Z_{\SW}$ is the contributions from 
the strong-coupling singularities of the moduli space 
$\calU$ (which is called the ``$u$-plane'' by abuse of 
the terminology for the $SU(2)$ and $SO(3)$ gauge groups).  
The second piece $Z_u$ is called the $u$-plane integral, 
which is absent if $b_2^{+}(X) > 1$.  This is the 
contributions from the whole moduli space $\calU$ 

According to Moore and Witten \cite{bib:Mo-Wi} 
(for $SU(2), SO(3)$ gauge groups) and Mari\~no and Moore 
\cite{bib:Ma-Mo} (for other gauge groups), the $u$-plane 
integral for the above Donaldson-Witten function $Z_{\DW}$ 
can be written 
\beqn
    Z_u = \int_\calU da d\abar 
          A^\chi B^\sigma \exp(U + S^2 T) \Psi, 
\eeqn
where $\chi$ and $\sigma$ are the Euler number and 
the signature of $X$; $A$ and $B$ are modular forms 
on $\calU$; $U$ is a contribution from $\calO(P)$ only; 
$T$ is a ``contact term'' which is induced by the 
intersection of the 2-cycle $S$ with itself,  
accordingly multiplied by the self-intersection 
number $S^2$; $\Psi$ is a lattice sum collecting 
the contributions of abelianized  gauge fileds 
and other fermionic degrees of freedom.

\subsection{Blowup formula and tau function} 

Mari\~no and Moore \cite{bib:Ma-Mo} pointed out that the 
blowup formula of the $u$-plane integral contains a factor 
that can be interpreted as a special ``tau function'' of 
the affine Toda system (or, more precisely, an underlying 
integrable hierarchy).  It should however be noted that 
this is the interpretation in the case of the $SU(N+1)$ 
gauge group.  

The blowup formula connects the manifold $X$ and 
its blowup $\Xtilde$ at a point $Q$.  $X$ is assumed 
to be a complex algebraic surface (e.g., $\bbC\bbP^2$, 
$\bbC\bbP^1 \times \bbC\bbP^1$, del Pezzo surfaces, etc.).  
Let $B$ denote the exceptional divisor (i.e., inverse 
image of $Q$) in $\Xtilde$, and consider the following 
Donaldson-Witten function of $\Xtilde$: 
\beqn
    \Ztilde_{DW} = 
    \Bigl< \exp\Bigl(tI(B) + I(S) + \calO(P)\Bigr)\Bigr>. 
\eeqn
Note that this path integral is over the fields on 
$\Xtilde$; the pull-back of $I(S)$ and $\calO(P)$ to 
$\Xtilde$ are denoted by the same notations.  The new 
observable $I(B)$ with support on $B$ is inserted with 
the coupling constant $t$.  The blowup formula then 
shows that the integrand of the $u$-plane integral 
for $\Ztilde_{\DW}$ is obtained by replacing 
\beqn
    e^U \ \to \ 
    e^U \frac{\alpha}{\beta} 
        \det\left(\frac{\rd u_k}{\rd a_j}\right)^{1/2}
        \Delta^{-1/8} 
        e^{-t^2T} \Theta_{\gamma,\delta}\Bigl(
          \frac{t}{2\pi}\vec{V} \mid \calP\Bigr)
    \label{eq:blowup-factor}
\eeqn
in the $u$-plane integral for $Z_{\DW}$. Various terms 
on the right hand side of this rule have the following 
meaning:  $\alpha$ and $\beta$ are some numerical constants; 
$\Delta$ is the discriminant of the family of the curves 
$C$ over $\calU$; $\Theta_{\gamma,\delta}(Z \mid \calP)$ 
is the ordinary $N$-dimensional theta function with 
characteristic $(\gamma,\delta)$ and period matrix 
$\calP$; $\vec{V}$ is the gradient vector 
\beqn
    \vec{V} = \left( \frac{\rd V}{\rd a_j} \right)
\eeqn
of the gauge invariant potential $V$ from which the 
integrand $G^2 V$ of $I(S)$ and $I(B)$ are constructed
(e.g., $V = \Tr \phi^2$ in the aforementioned usual setup 
of topological gauge theories).  For the $SU(N+1)$ 
Seiberg-Witten curve, the matrix $\calP$ is the period 
matrix of $\Jac(C)$; for the other classical gauge groups, 
the period matrix of $\Prym(C/C_2)$ (or $\Jac(C')$) appears.  
The characteristic $(\gamma,\delta)$ is determined by 
the physical setup; typically, $\gamma = (0,\cdots,0)$ 
and $\delta = (1/2,\cdots,1/2)$. 

It is the product of the last two terms in 
(\ref{eq:blowup-factor}) that Mari\~no and Moore, 
in the case of the $SU(N+1)$ gauge group, identified 
to be the tau function of the affine Toda system: 
\beqn
    \tau_{\gamma,\delta}(t) = 
      e^{-t^2T} \Theta_{\gamma,\delta}\Bigl(
      \frac{t}{2\pi}\vec{V} \mid \calP\Bigr). 
\eeqn
Thus, the coupling constant $t$ plays the role of 
a time variable in the $A^{(1)}_N$ Toda system.  
For the other classical gauge groups, however, the 
relation to the affine Toda systems is slightly 
more complicated, as we discuss later on.

\subsection{Multi-time tau function as blowup factor}

Our previous paper \cite{bib:Ta-u-plane} proposes 
a ``multi-time'' analogue of the single-time tau function 
$\tau_{\gamma,\delta}(t)$ above.  In the present setup 
including all classical gauge groups, the multi-time 
analogue can be written 
\beqn
   \tau_{\gamma,\delta}(t_1,t_2,\cdots) = 
     \exp\Bigl( \frac{1}{2}\sum_{m,n\ge 1}q_{mn}t_mt_n \Bigr)
     \Theta_{\gamma,\delta}\Bigl(\sum_{n\ge 1}t_n\vec{V}^{(n)} 
       \mid \calP\Bigr). 
\eeqn
The coefficients $q_{mn}$ of the Gaussian factor 
and the components of the directional vectors 
$\vec{V}^{(n)} = \Bigl(V^{(n)}_j\Bigr)$ are written 
in terms of period integrals of the meromorphic 
differentials $d\Omega_n$ that we have considered: 
\beqn
    q_{mn} &=& 
      - \frac{1}{2\pi i}\oint_{P_\infty} f_n(z) d\Omega_m 
      - \frac{1}{2\pi i}\oint_{Q_\infty} g_n(z) d\Omega_m, 
    \nonumber \\
    V^{(n)}_j &=& 
      - \frac{1}{2\pi i}\oint_{P_\infty} f_n(z) d\omega_j
      - \frac{1}{2\pi i}\oint_{Q_\infty} g_n(z) d\omega_j. 
\eeqn
By Riemann's bilinear relation, $V^{(n)}_j$ can also be 
written 
\beqn
    V^{(n)}_j = \frac{1}{2\pi i} \oint_{B_j} d\Omega_n. 
\eeqn
Our proposal in the previous paper (for the $SU(N+1)$ 
topological gauge theory) is to interpret this tau function 
as the counterpart of Mari\~no and Moore's blowup factor 
$\tau_{\gamma,\delta}(t)$ for the Donaldson-Witten function 
\beqn
    \Ztilde_{DW} = 
    \Bigl< \exp\Bigl(\sum_{n\ge 1} t_n I_n(B) 
         + I(S) + \calO(P)\Bigr)\Bigr>. 
\eeqn
with many 2-cycles observables $I_n(B)$ inserted.  
The observables $I_n(B)$ are of the form 
\beqn
    I_n(B) = \const \int_B G^2 V^{(n)}, 
\eeqn
and the directional vector $\vec{V}^{(n)}$ is the gradient 
of the gauge invariant potential $V^{(n)}$, 
\beqn
    \vec{V}^{(n)} = \Bigl( \frac{\rd V^{(n)}}{\rd a_j} \Bigr). 
\eeqn
This leads to the identification of the coefficients 
$q_{mn}$ as the ``contact terms'' $C\Bigl(V^{(m)},V^{(n)}\Bigr)$ 
of higher Casimir observables in the sense of Losev et al. 
\cite{bib:Lo-Ne-Sh}.  

Strong evidence supporting our proposal is that the above 
multi-time tau function has a good modular property under 
the symplectic transformations 
\beqn
    \begin{array}{ll}
    B_j \to \calA_{jk} B_k + \calB_{jk} A_k, \\
    A_j \to \calC_{jk} B_k + \calD_{jk} A_k, 
    \end{array}
    \quad 
    \left( \begin{array}{ll}
      \calA & \calB \\
      \calC & \calD
    \end{array} \right)  \in Sp(2N, \bbZ).  
    \label{eq:Sp-on-cycles}
\eeqn
of the cycles $A_j,B_j$.  This is crucial for ensuring 
the correct modular property of the integrand of the 
$u$-plane integral.  Under this symplectic transformation, 
indeed, the tau function $\tau_{\gamma,\delta}$ transforms as 
\beqn
    \tau_{\gamma,\delta}(t_1,t_2,\cdots) \to 
    \epsilon \det(\calC \calP + \calD)^{1/2} 
    \tau_{\gamma',\delta'}(t_1,t_2,\cdots), 
\eeqn
where $\epsilon$ (an 8th root of unity), $\gamma'$ and 
$\delta'$ are determined by the $Sp(2N,\bbZ)$ matrix. 
This fact can be confirmed in the same way as the 
proof in the case of the $SU(N+1)$ topological gauge 
theory \cite{bib:Ta-u-plane}.  It should be also 
mentioned that this modular property of the tau 
function has been known for years 
\cite{bib:AG-Mo-Va,bib:KNTY}.  Now the point is 
that the modular property of $\tau_{\gamma,\delta}$ 
is {\it independent of} $t_1,t_2,\cdots$.  
In particular, the modular invariance of the 
$u$-plane integral in Mari\~no and Moore's setup 
at $t_1 = t$ and $t_n = 0$ ($n > 1$) is retained 
even if the higher coupling constants $t_n$ are 
turned on.

\subsection{KdV again} 

What is new in the case of the orthogonal and 
symplectic groups is that the theta function in 
the tau function is {\it not} a theta function on 
$\Jac(C)$; this is a theta function on the Prym 
variety $\Prym(C/C_2)$ or, up to an isogeny, 
the Jacobi variety $\Jac(C')$ of the quotient 
curve $C'$.  As remarked in the previous sections, 
the quotient curve $C'$ and the meromorphic 
differentials $d\Omega_n$ are of the KdV type. 
The tau function $\tau_{\gamma,\delta}$ for 
the orthogonal and symplectic gauge groups 
thus turns out to be a tau function of the 
KdV hierarchy (which is a special case of 
the algebro-geometric tau functions of the 
KP hierarchy\cite{bib:Mulase,bib:Shiota}) 
rather than of the affine Toda system. 

This conclusion might cause confusion, but here is 
no contradiction.  It is well known that the affine 
Toda system can be mapped to linear flows on the 
Prym variety \cite{bib:Ad-vM}.  Apart from the case 
of $A^{(1)}_N$, however, this does {\it not} imply 
that its solutions and tau functions can be written 
in terms of theta functions on the Prym variety.  
Actually, the flows are first linearized on the Jacobi 
variety $\Jac(C)$ of the affine Toda spectral curve 
itself, and then shown to be confined to a subspace 
that is {\it parallel} (but {\it not identical}) to 
the Prym variety embedded therein.  All that one can 
expect is, accordingly, an expression in terms of 
theta functions on $\Jac(C)$.  (Surprisingly, however, 
very few is known about such an {\it explicit} 
expression of solutions of the affine Toda systems 
other than the $A^{(1)}_N$ and $C^{(1)}_N$.)   
Thus, our tau function $\tau_{\gamma,\delta}$ is 
something different from tau functions of the 
affine Toda systems, and our interpretation is 
that it is a special tau function of the KdV type.

\subsection{Whitham deformations and prepotential} 

Let us now turn on the Whitham deformations with 
``slow variables'' $T_n$.  The Seiberg-Witten 
prepotential $\calF$ is also deformed and becomes 
a function $\calF(\vec{a},\vec{T})$ of both 
$\vec{a} = (a_1,\cdots,a_N)$ and 
$\vec{T} = (T_1,T_2,\cdots)$.  More precisely, 
$\calF$ is defined by (\ref{eq:define-F}) for 
all cases considered in the preceding sections.  
As one can immediately  see by comparing these 
equations with the definition of $q_{mn}$, 
$V^{(n)}_j$ and $\calP_{jk}$ in the form of 
period integrals, these fundamental quantities 
in our interpretation of the blowup formula 
can be written as second order derivatives of 
the prepotential:
\beqn
    V_{jn} = \frac{1}{2 \pi i}\frac{\rd^2 \calF}{\rd a_j \rd T_n}, 
    \quad 
    q_{mn} = \frac{1}{2 \pi i}\frac{\rd^2 \calF}{\rd T_m \rd T_n}, 
    \quad 
    \calP_{jk} = \frac{\rd^2 \calF}{\rd a_j \rd a_k}. 
\eeqn
In particular, the potential $V^{(n)}$ of the observables 
$I_n(B)$ turn out to be first order derivatives of $\calF$: 
\beqn
    V^{(n)} = \frac{1}{2\pi i}\frac{\rd \calF}{\rd T_n}. 
\eeqn
Of course the $a_j$-derivatives give the dual special
coordinates $a^D_j = \oint_{B_j} dS$: 
\beqn
    a^D_j = \frac{\rd \calF}{\rd a_j}. 
\eeqn
Thus the prepotential $\calF = \calF(\vec{a},\vec{T})$ 
in the Whitham deformations, too, is a kind of 
``generating function''.  

Unlike the ``fast variables'' $t_n$, the role of 
the ``slow variables'' $T_n$ is to deform the 
period map $\vec{u} \mapsto \vec{a}$ that connects 
the $u$-space and the $a$-space.  Presumably, this will 
be a kind of deformations of ``background geometry'' 
in the sense of string theory, but the  precise 
meaning is still beyond our scope.

%%%%%%%%%%%%%%%%%%%%%%%%%%%%%%%%%%%%%%%%%%%%%%%%%%%%%%%%%%%%%%%%%%%%

\section{Discussions}
\setcounter{equation}{0}

We have seen that the construction of Whitham deformations 
by Gorsky et al. \cite{bib:GMMM} can be extended to the 
Seiberg-Witten curves of all the classical gauge groups 
and some other complex algebraic curves.  Although the 
construction is based on the somewhat special form 
(\ref{eq:SW-curve}) of the curves, the only requirement 
seems to be that  $W(x)$ be a rational function with 
a polynomial leading part. Actually, we have obtained 
partial evidence that an irrational superpotential might 
be allowed for at least in some special cases. 

We have also extended our proposal in the previous 
paper \cite{bib:Ta-u-plane} on the $u$-plane integral 
of the $SU(N+1)$ topological gauge theory to all 
other classical gauge groups.  A byproduct of the 
construction of Whitham deformations is to determine 
which flows of the underlying integrable hierarchy 
(the Toda hierarchy for the case of $SU(N+1)$ and 
the KdV hierarchy for the other classical gauge groups) 
should be extracted; appropriate flows are those generated 
by the meromorphic differentials $d\Omega_n$ that arise 
in the construction of Whitham deformations.  This enables 
us to express the relevant quantities $q_{mn}$ etc. as 
derivatives of the prepotential $\calF$.  

Let us conclude this paper with discussions on possible 
extensions and implications of these results.   

\subsection*{Non-hyperelliptic curves}

The first nontrivial step beyond rational superpotentials 
is irrational superpotentials of the form 
\beqn
    W(x) = R_1(x) + R_2(x) \sqrt{R_3(x)}, 
\eeqn
where $R_i(x)$'s are polynomial or rational functions of $x$.  
This means that the curve $C$ is no longer hyperelliptic, 
but can be written in a special quartic polynomial in $z$ 
with rational coefficients.  Well known examples of 
curves of this form  \cite{bib:Ma-Wa,bib:Le-Wa-E6} 
are the Seiberg-Witten curves of $SU(5)$ ($A_4$) in 
the 10-dimensional anti-symmetric representation, 
$E_6$ in the 27-dimensional minimal representation, 
and $G_2$ in the 7-dimensional minimal representation. 

An obvious difficulty is that $W(x)$ itself is multi-valued, 
so that the fractional powers of $W(x)$ requires a more 
careful treatment.  This difficulty, however, might be 
easily overcome, because the same fractional powers are 
used in the work of Eguchi and Yang \cite{bib:Eg-Ya-E6} 
on the topological Landau-Ginzburg of the $E_6$ singularity. 
This work also predicts an interesting phenomena if 
the fractional power construction really works for the 
case of $E_6$. Namely, as they observed in the topological 
Landau-Ginzburg theory, the admissible Whitham deformations 
will be limited to those associated with the fractional 
powers $W(x)^{n/12}$ with 
\beqn
    n \equiv 1,4,5,7,8,11 \ \mod 12. 
\eeqn
The numbers on the right hand side are the exponents of 
$E_6$. 

For more general cases, however, an entirely new approach 
will be necessary.  For instance, Witten's M5-brane 
construction \cite{bib:Wi-M5} yields a non-hyperelliptic 
curve of the form 
\beqn
    z^{k+1} + g_1(x) z^k + \cdots + g_k(x)z + 1 = 0 
\eeqn
for the $\calN = 2$ SUSY gauge theory (coupled to 
bifundamental matters) with the product gauge group 
$SU(N_1) \times \cdots \times SU(N_k)$. 
The Seiberg-Witten differential is given by 
\beqn
    dS_{\SW} = x \frac{dz}{z}. 
\eeqn
A natural ansatz for the meromorphic differentials 
$d\Omegahat_n$ of Whitham deformations is to seek for them 
in the form 
\beqn
    d\Omegahat_n = R_n(x,z) \frac{dz}{z}, \quad 
    R_n(x,z) = \mbox{polynomial}. 
\eeqn
We do not know how to construct the polynomials $R_n(x,z)$.  
The problem becomes even harder for the elliptic models 
of M5-branes.

\subsection*{Relation to topological Landau-Ginzburg theories}

The fractional power construction strongly suggests a 
direct link with topological Landan-Ginzburg theories of 
A-D-E singularities coupled to gravity or, equivalently, 
$d < 1$ topological strings \cite{bib:DVV,bib:Bl-Va}. 
The relation between the Seiberg-Witten theory and 
$d < 1$ topological strings has been studied from 
several aspects, such as the WDVV equations 
\cite{bib:Ma-etal-WDVV,bib:Be-Ma-WDVV}, 
flat coordinates and Gauss-Manin systems 
\cite{bib:It-Ya-GM,bib:It-Xi-Ya-GM}, 
etc.  Of course the very notion of prepotentials 
itself is a bridge connecting the two worlds.  
In the $d < 1$ topological strings, the role of 
the Whitham equations is played by the dispersionless 
limit of integrable hierarchies 
\cite{bib:Kr-LG,bib:Du-LG,bib:Ta-Ta-review}. 
The fractional powers of the superpotential are 
fundamental building blocks of the Lax representation 
therein. Nevertheless, the emergence of fractional 
powers in the Seiberg-Witten theory is quite surprising.  

An interesting outcome of our Whitham deformations 
is that they have an exotic limit as $\mu \to 0$. 
In this limit, the Seiberg-Witten curve reduces 
to the {\it rational curve} 
\beqn
    z = W(x), 
    \label{eq:rat-SW-curve}
\eeqn
and the Seiberg-Witten differential turns into 
the rational differential 
\beqn
    dS_{\SW} = x \frac{W'(x)dx}{W(x)}.
\eeqn
As we shall show below, the Whitham equations, 
too, have a well defined limit. Furthremore, 
these differential equations are similar, but 
{\it not identical}, to the following counterpart 
in $d < 1$ topological strings 
\cite{bib:Kr-LG,bib:Du-LG,bib:Ta-Ta-review}: 
\beqn
    \left.\frac{\rd}{\rd T_n}
      \sum_{m\ge 1}T_m R_m(x)\right|_{W(x)=\const} 
    = R_n(x). 
\eeqn

This difference stems from the difference of the 
two theories as Landau-Ginzburg models.  Namely, 
whereas the Whitham deformations at $\mu = 0$ is 
still related to a {\it curve} defined by 
(\ref{eq:rat-SW-curve}),  the Landau-Ginzburg 
description of $d < 1$ topological strings is 
based on a {\it 0-dimensional manifold} 
defined by the equation 
\beqn
    W(x) = 0.  
\eeqn

Now, let us present the Whitham equations at $\mu = 0$. 
For simplicity, we consider the case of the $SU(N+1)$ 
curve where $W(x) = P(x)$; the other case can be treated 
similarly.  Suppose, as usual, that the cycles $A_j$ 
are chosen to encircle the cuts between two neighboring 
roots $e_j^{\pm}$ of $P(x) - 4\mu^2$. As $\mu \to 0$,  
the $j$-th cuts shrink to a point at the $j$-th root 
$e_j$ of $P(x) = \prod_{j=1}^{N+1}(x - e_j)$.  
The period integrals $a_j = \oint_{A_j} dS$ then 
reduce to residue integrals, which can be readily calculated: 
\beqn
    a_j = 2 \pi i \sum_{n\ge 1} T_n R_n(e_j) 
    \quad (n = 1,\cdots,N). 
\eeqn
This defines a map $\vec{u} \mapsto \vec{a}$ from the 
$\vec{u}$-space to the $\vec{a}$-space with deformation 
parameters $T_n$, and this map is invertible if 
$\vec{T}$ is close to $(1,0,0,\cdots)$. 
(Note that the $N+1$-th root $e_{N+1}$ of $P(x)$ is not 
independent; the roots of $P(x)$ obeys the constraint 
$\sum_{j=1}^{N+1} e_j = 0$.) The inverse map determines, 
as in the case with $\mu \not= 0$, a family of deformations 
of the rational curve $z = P(x)$.  Under these deformations, 
the following equations can be eventually derived: 
\beqn
    \left.\frac{\rd}{\rd T_n}
      \sum_{m\ge 1}T_m R_m(x)\right|_{P(x)=\const} 
    &=& 
    R_n(x) - \sum_{k=1}^{N+1}
      \frac{R_n(e_k)P(x)}{(x - e_k)P'(x)}
    \nonumber \\
    \left.\frac{\rd}{\rd a_j}
      \sum_{m\ge 1}T_m R_m(x)\right|_{P(x)=\const}
    &=& 
    \left(\frac{1}{x - e_j} - \frac{1}{x - e_{N+1}}\right)  
      \frac{P(x)}{2\pi iP'(x)} 
\eeqn
We omit the proof of these equations, but the following 
comment would be enough for understanding:  These equations 
can be derived from the Whitham equations (\ref{eq:Whitham}) 
if $d\Omega_n$ and $d\omega_j$ are interpreted as follows: 
\beqn
    d\Omega_n &=& 
      R_n(x)\frac{P'(x)dx}{P(x)} 
      - \sum_{j=1}^N 2\pi i R_n(e_j) d\omega_j, 
    \nonumber \\
    d\omega_j &=& 
      \left(\frac{1}{x - e_j} - \frac{1}{x - e_{N+1}}
      \right) \frac{dx}{2 \pi i}. 
\eeqn
In fact, they give a correct limit, as $\mu \to 0$, 
of the differentials on the $\mu \not= 0$ curve.

\subsection*{Acknowledgements}

I am grateful to Toshio Nakatsu for useful discussions,
and to Yuji Shimizu for bibliographic information 
on Prym varieties.  This work is partly supported by the 
Grant-in-Aid for Scientific Research (No. 10640165) from 
the Ministry of Education, Science and Culture. 

%%%%%%%%%%%%%%%%%%%%%%%%%%%%%%%%%%%%%%%%%%%%%%%%%%%%%%%%%%%%%%%%%%%%

\appendix
\renewcommand{\theequation}{\Alph{section}.\arabic{equation}}

\section{Spectral Curves of Affine Toda Systems}
\setcounter{equation}{0}

Here we present a list of the Toda spectral curves 
\beqn
    \det\Big(L(z) - xI\Bigr) = 0 
\eeqn
used in the text, along with the $L$-matrices $L(z)$.  
The $L$-matrices are realized in a representation of 
minimal dimensions.  The $L$-matrices for $A^{(1)}_N$, 
$B^{(1)}_N$, $C^{(1)}_N$ and $D^{(1)}_N$ are borrowed 
from the work of Adler and van Moerbeke \cite{bib:Ad-vM}.  
The other cases associated with the twisted affine 
algebras $A^{(2)}_N$ and $D^{(2)}_N$ are derived by 
the following ``folding'' procedure: 
\beqn
    D^{(1)}_{2N} \mapsto A^{(2)}_{2N-1}, \quad 
    D^{(1)}_{N+2} \mapsto D^{(2)}_{N+1}, \quad 
    D^{(1)}_{2N+2} \mapsto A^{(2)}_{2N}. 
\eeqn

The $L$-matrices in the following list have a 
``symmetric'' form, i.e., $L(z)^T = L(z^{-1})$, 
as opposed to the $L$-matrices of Martinec and 
Warner \cite{bib:Ma-Wa}. Accordingly, the actual 
form of the spectral curves becomes 
\beqn
    \mu(z + z^{-1}) = W(x). 
\eeqn
upon removing an overall constant or a non-dynamical 
factor (the factor $x$ the case of $B^{(1)}_N$). 
This curve, however, can be readily converted to the form 
of Marninec and Warner by recalling $z \to z/\mu$.  

Some comments on the notations in the list are in order.  
$a_j$ and $b_j$ ($j = 1,\cdots,N$) are related to the 
canonical variables $q$ and $p$  (both in the Cartan 
subalgebra of the classical part of the affine algebra) 
of the affine Toda system as follows: 
\beqn
    a_j = c_j g e^{\alpha_j \cdot q}, \quad 
    b_j = p_j = h_j\cdot p. 
\eeqn
Here $\alpha_j$ ($j = 1,\cdots,N$) are the simple roots 
and $\alpha_0$ is the affine root.  (The null root is 
ignored.)   $c_j$ are numerical constants related 
to the root system, and $g$ the coupling constant.   
(The $a_j$'s should not be confused with the special 
coordinates $a_j$'s in the Seiberg-Witten theory.)  
$E_{jk}$ denotes the matrix with the only non-vanishing 
elements equal to $1$ at $(j,k)$: 
\beqn
    (E_{jk})_{mn} = \delta_{jm}\delta_{kn}. 
\eeqn

\begin{enumerate}
\item $A^{(1)}_N$: The $L$-matrix is $(N+1) \times (N+1)$. 
\beqn
 && \begin{array}{lll}
    L(x) 
    &=& \sum_{j=1}^{N+1} a_j E_{j,j+1} + a_0 zE_{N1} 
        + \sum_{j=1}^{N+1} b_j E_{jj} 
        + \sum_{j=1}^{N+1} a_j E_{j+1,j} \\
    &&  + a_0 z^{-1}E_{1N}. 
    \end{array} 
    \nonumber \\
 && \det\Bigl(L(z) - xI\Bigr) = 
      (-1)^N A(z + z^{-1}) + (-1)^{N+1} P(x), 
    \nonumber \\
 && A = a_0 a_1 \cdots a_N, \quad 
    \mu = A. 
    \nonumber 
\eeqn

\item $B^{(1)}_N$: The $L$-matrix is $(2N+1) \times  (2N+1)$. 
\beqn
 && \begin{array}{lll}
    L(x) 
    &=& \sum_{j=1}^N a_j (E_{j,j+1} - E_{2N+1-j,2N+2-j}) 
        + a_0 (zE_{2N+1,2} - zE_{2N,1})  \\
    &&  + \sum_{j=1}^N b_j (E_{jj} - E_{2N+2-j,2N+2-j}) \\
    &&  + \sum_{j=1}^N a_j (E_{j+1,j} - E_{2N+2-j,2N+1-j}) 
        + a_0 (z^{-1}E_{2,2N+1} - z^{-1}E_{1,2N}). 
    \end{array}
    \nonumber \\
 && \det\Bigl(L(z) - xI\Bigr) = 
      x \Bigl( 2(-1)^N A(z + z^{-1}) - Q(x^2) \Bigr). 
    \nonumber \\
 && A = a_0 a_1 a_2^2 \cdots a_N^2, \quad 
    \mu = 2 (-1)^N A. 
    \nonumber 
\eeqn
\item $C^{(1)}_N$: The $L$-matrix is $2N \times 2N$. 
\beqn
 && \begin{array}{lll}
    L(x) 
    &=& \sum_{j=1}^{N-1} a_j (E_{j,j+1} - E_{2N-j,2N+1-j}) 
        + a_N E_{N,N+1} + a_0 zE_{2N,1} \\
    &&  + \sum_{j=1}^N b_j (E_{jj} - E_{2N+1-j,2N+1-j}) \\
    &&  + \sum_{j=1}^{N-1} a_j (E_{j+1,j} - E_{2N+1-j,2N-j}) 
        + a_N E_{N+1,N} + a_0 z^{-1} E_{1,2N}.
    \end{array}
    \nonumber \\
 && \det\Bigl(L(z) - xI\Bigr) = 
      (-1)^N A (z + z^{-1}) + Q(x^2). 
    \nonumber \\
 && A = a_0 a_1^2 \cdots a_{N-1}^2 a_N, \quad 
    \mu = - (-1)^N A. 
    \nonumber 
\eeqn
\item $D^{(1)}_N$: The $L$-matrix is $2N \times 2N$. 
\beqn
 && \begin{array}{lll}
    L(z) 
    &=& \sum_{j=1}^{N-1} a_j (E_{j,j+1} - E_{2N-j,2N+1-j}) 
        + a_N (E_{N,N+2} - E_{N-1,N+1}) \\
    &&  + a_0 (zE_{2N,2} - zE_{2N-1,1}) 
        + \sum_{j=1}^N b_j (E_{jj} - E_{2N+1-j,2N+1-j}) \\
    &&  + \sum_{j=1}^{N-1} a_j (E_{j+1,j} - E_{2N+1-j,2N-j}) 
        + a_N (E_{N+2,N} - E_{N+1,N-1}) \\
    &&   + a_0 (z^{-1} E_{2,2N} - z^{-1}E_{1,2N-1}). 
    \end{array} 
    \nonumber \\
 && \det\Bigl(L(z) - xI\Bigr) = 
      - 4(-1)^N A x^2 (z + z^{-1}) + Q(x^2). 
    \nonumber \\
 && A = a_0 a_1 a_2^2 \cdots a_{N-2}^2 a_{N-1} a_N, \quad 
    \mu = 4 (-1)^N A. 
    \nonumber 
\eeqn
\item $A^{(2)}_{2N-1}$: The $L$-matrix is $2N \times 2N$. 
\beqn
 && \begin{array}{lll}
    L(z) 
    &=& \sum_{j=1}^{N-1} a_j (E_{j,j+1} - E_{2N-j,2N+1-j}) 
        + a_N E_{N,N+1} \\
    &&  + a_0 (zE_{2N,2} - zE_{2N-1,1}) 
        + \sum_{j=1}^N b_j (E_{jj} - E_{2N+1-j,2N+1-j}) \\
    &&  + \sum_{j=1}^{N-1} a_j (E_{j+1,j} - E_{2N+1-j,2N-j}) 
        + a_N E_{N+1,N} \\
    &&  + a_0 (z^{-1}E_{2,2N} - z^{-1}E_{1,2N-1}). 
    \end{array}
    \nonumber \\
 && \det\Bigl(L(z) -xI\Bigr) = 
      2(-1)^N A x (z + z^{-1}) + Q(x^2). 
    \nonumber \\
 && A = a_0 a_1 a_2^2 \cdots a_{N-1}^2 a_N, \quad 
    \mu = - 2(-1)^N A. 
    \nonumber 
\eeqn
\item $D^{(2)}_{N+1}$: The $L$-matrix is $(2N+2) \times (2N+2)$. 
\beqn
 && \begin{array}{lll}
    L(z) 
    &=& \sum_{j=1}^N a_j (E_{j+1,j+2} - E_{2N+2-j,2N+3-j}) 
        + a_0 (zE_{2N+2,1} - E_{12}) \\
    &&  + \sum_{j=1}^N b_j (E_{j+1,j+1} - E_{2N+3-j,2N+3-j}) \\
    &&  + \sum_{j=1}^N a_j (E_{j+2,j+1} - E_{2N+3-j,2N+2-j}) 
        + a_0 (z^{-1}E_{1,2N+2} - E_{21}). 
    \end{array}
    \nonumber \\
 && \det\Bigl(L(z) - xI\Bigr) = 
      (-1)^N A (z + z^{-1} - 2) + x^2 Q(x^2). 
    \nonumber \\
 && A = a_0^2 a_1^2 \cdots a_N^2, \quad 
    \mu = - (-1)^N A. 
    \nonumber 
\eeqn
\item $A^{(2)}_{2N}$: The $L$-matrix is $(2N+1) \times (2N+1)$. 
\beqn
 && \begin{array}{lll}
    L(z) 
    &=& \sum_{j=1}^{N-1} a_j (E_{j+1,j+2} - E_{2N+1-j,2N+2-j}) 
        + a_N E_{N+1,N+2} \\
    &&  + a_0 (zE_{2N+1,1} - E_{12}) 
        + \sum_{j=0}^N b_j (E_{j+1,j+1} - E_{2N+2-j,2N+2-j}) \\
    &&  + \sum_{j=1}^{N-1} a_j (E_{j+2,j+1} - E_{2N+2-j,2N+1-j}) 
        + a_N E_{N+2,N+1} \\
    &&  + a_0 (z^{-1}E_{1,2N+1} - E_{21}). 
    \end{array} 
    \nonumber \\
 && \det\Bigl(L(z) - xI\Bigr) = 
      (-1)^N A (z + z^{-1}) + x Q(x^2). 
    \nonumber \\
 && A = a_0^2 a_1^2 \cdots a_{N-1}^2 a_N, \quad 
    \mu = - (-1)^N A. 
    \nonumber 
\eeqn
\end{enumerate}

%%%%%%%%%%%%%%%%%%%%%%%%%%%%%%%%%%%%%%%%%%%%%%%%%%%%%%%%%%%%%%%%%%%%

\end{document}